\documentclass[12pt]{article}
\usepackage[centertags]{amsmath}
\usepackage{amsfonts}
\usepackage{amssymb}
\usepackage{amsthm} 
\usepackage{newlfont}
\usepackage{epsfig}
\usepackage{amscd}
\usepackage{graphicx}
\usepackage{epsfig}
\usepackage{amscd}
\usepackage{color}

\newcommand{\beq}{\begin{equation}}
\newcommand{\eeq}{\end{equation}}
\newcommand{\ba}{\begin{array}}
\newcommand{\ea}{\end{array}}
\newcommand{\bea}{\begin{eqnarray}}
\newcommand{\eea}{\end{eqnarray}}
\begin{document}

\begin{center}

{\large \sc \bf {The minimal entanglement of bipartite decompositions as a
witness  of  strong entanglement in a quantum system.}}

\vskip 15pt

{\large A.I.~Zenchuk }

\vskip 8pt

{\it Institute of Problems of Chemical Physics, Russian Academy of Sciences,
Chernogolovka, Moscow reg., 142432, Russia, e-mail:   zenchuk@itp.ac.ru } 
\today
\end{center}

\begin{abstract}
We  {characterize the  multipartite entanglement in a quantum system by the 
quantity} which vanishes if only the quantum
system may be decomposed into two weakly entangled subsystems, unlike 
measures of 
 multipartite entanglement introduced before. We refer to this {quantity} as the minimal
 entanglement of bipartite decompositions (MEBD).  
Big MEBD means that the system may not be decomposed into two weakly entangled
subsystems. MEBD allows one to define, for instance,  whether the given quantum
system may be a candidate for a quantum register, where the above decomposition
is undesirable. 
 A method of lower estimation of MEBD is represented. Examples of big MEBD in
spin-1/2 chains governed by the $H_{dz}$ Hamiltonian in the strong external
magnetic field  are given.
\end{abstract}

\section{Introduction}

The problem of  strong entanglement in spin systems is very important in view of
 development of quantum computation. 
{It is acknowledged that the quantum correlations are needed in order to organize the coherent  manipulations by different bits in quantum circuit. This is a basic resource of quantum computation providing advantages of quantum circuits in comparison with their classical counterparts \cite{NC}. In particular, quantum correlations are responsible for  the speedup of quantum algorithms.  It is hard to create  large  strongly entangled systems. On the contrary,  a few-qubit quantum registers have been constructed  using different physical basis. For instance, there are registers using systems of coupled trapped ions \cite{Lea}, registers using superconducting charge qubits
\cite{YPANT} and 
 neutral atom quantum registers \cite{SDKMRM}.
  These small registers can be organized in large scale quantum circuits \cite{DB,LBBKK,ODH}.}
Review of different measures of both bipartite and multipartite entanglements is
given, for instance, in \cite{AFOV,HHHH}.
 { Especially one has to mention the Wootters criterion which is well elaborated for calculation  of  entanglement among two  nodes in
a  spin-1/2 system \cite{HW,W}}. The entanglement between two subsystems consisting
of more  then one node  may be effectively described by the positive partial
transpose (PPT) criterion \cite{P,VW} introducing so-called double negativity as
a measure of
entanglement. {This measure is applicable to both pure and mixed states.}
An important concept is so-called entanglement of formation \cite{BDSW}
describing the entanglement between two subsystems in a mixed state. 

This paper is devoted to the problem of multipartite entanglement.
Whereas the bipartite entanglement is studied in many details, the multipartite
entanglement is more cumbersome \cite{VPRK,WG} and it has  not been completely
understood yet. 
Of course, considering the problem of strong multipartite entanglement in a
quantum system, it would be reasonable to construct such strongly entangled 
system, { where any two nodes} are strongly entangled simultaneously.
However, this requirement seems to be  too tough. 
Therefore  we assume that the strong node-to-node entanglement between any two
nodes is not necessary in order to entangle all nodes in the quantum system.
Instead of this, we state that the system becomes strongly entangled at some time
moment $t_0$ if any its subsystem is strongly entangled with the rest of the
system at this time moment. 

{
We suggest  a method to pick out such quantum systems which may not be decomposed into two weakly entangled subsystems. It is shown that such systems may be simply realized using spin chains. Entanglement considered here is  some kind of "collective" entanglement because not all node-to-node entanglements in such systems are big. We assume that such "collective" entanglement may provide advantages of quantum devices in comparison with they classical counterparts.}

Following the above two paragraphs, we introduce so-called minimal entanglement of
bipartite decompositions (MEBD)  of the quantum system as a witness of strong
entanglement, or as the measure of the above mentioned "collective" entanglement. The basic feature of MEBD is that its  measure  vanishes if only
there is at least one possibility to decompose  the given  quantum system into
two weakly entangled subsystems. Thus, MEBD helps us to test, for instance,
whether the given quantum system may be a candidate for a large volume  quantum register,
where such decomposition is undesirable {since one has to be able to  address  to all qbits  of register at ones with a single operation. Namely this property of quantum register is responsible for the exponential speedup of quantum algorithms in comparison with their classical counterparts}. In other words, 
if MEBD vanishes, then the system must be considered as two (or even more)
weakly entangled subsystems rather then a single entangled quantum system.
Emphasize that  big MEBD does not require significant entanglement between any
two particular nodes simultaneously. Instead of this, a big MEBD means that the
whole quantum system is entangled and there is no  decomposition of this system
into two un-entangled subsystems.

{
We consider the dynamics of  MEBD in homogeneous spin-1/2
chains governed by the $H_{dz}$ Hamiltonian in the strong external magnetic
field}. It will be demonstrated  that big MEBD is
achievable during the relatively short time interval 
(comparable with the end-to-end state transfer time interval $\tau$ along the properly adjusted inhomogeneous spin-1/2 chain,  $\tau\sim\pi$ \cite{CDEL,KS}) and this time interval { is
slightly increasing with the length of the spin chain}. 
It is also remarkable that  big MEBD is achievable in relatively simple models
such as  the homogeneous spin-1/2 chains with special initial conditions. {These special initial conditions are represented by the certain number of the initially excited spins, i.e. by
the  number of spins which are directed opposite to the external magnetic field initially.}
Due to these initial conditions  we are not forced to use
either the inhomogeneous chains \cite{VGIZ,GMT} or inhomogeneous magnetic field
\cite{DZ} to obtain big MEBD, unlike the node-to-node entanglement {in the spin chains with the single excited node} which has
been considered, for instance, in \cite{VGIZ,GMT,DZ}.

The definition of MEBD is introduced in   Sec.\ref{Section:def}.   An algorithm
allowing one to obtain a lower estimations of MEBD  is given in
Sec.\ref{Section:low}. This lower estimation becomes important for large
systems, where the calculation of MEBD following its definition is very
complicated. Examples of big MEBD in homogeneous spin-1/2 chains governed by the
$H_{dz}$ Hamiltonian in the strong external magnetic field are represented in
Sec.\ref{Section:Hdz}. 
{ We restrict ourselves to the short spin chains because
even multipartite entanglement in the short chains is not a simple phenomenon.
For instance, it is impossible 
 to organize the strong entanglement between any two nodes in the three node
spin-1/2 chain simultaneously, while big MEBD  is achievable in three-node and
larger spin chains.}
Basic results and conclusions are summarized in  
Sec.\ref{Section:conclusions}.

\section{The minimal entanglement of bipartite decompositions of a quantum
system}
\label{Section:def}


{ Hereafter we will characterize the entanglement  between two 
subsystems $a$ and $b$ of the $N$-node quantum  system $S_N=a\cup b$
($a\cap b=\emptyset$) in terms of the double
negativity ${\cal{N}}_{a,b}\equiv {\cal{N}}_{b,a}$ (PPT criterion \cite{P,VW}), where ${\cal{N}}_{a,b}$ is the
absolute value of the double sum of the negative eigenvalues of the matrix
$\rho^{T_a}$ (or $\rho^{T_b}$) which is the transposition of the 
 density matrix $\rho$ (associated with the system $S_N$)  
with respect to the subsystem $a$ (or $b$).  
More general, we will also use the bipartite double negativity ${\cal{N}}_{a_i,a_j}$ in the $L$-partite system $S_N$,
\begin{eqnarray}\label{multipartite} 
S_N=a_1\cup a_2 \cup\cdots\cup a_L,\;\;\;a_i\cap a_j =\emptyset, \;\;{\mbox{if}} \;\; i\neq j.
\end{eqnarray}
 In this case 
${\cal{N}}_{a_i,a_j}\equiv {\cal{N}}_{a_j,a_i}$ is the
absolute value of the double sum of the negative eigenvalues of the matrix
$\rho^{T_{a_j}}_{ij}$ (or $\rho^{T_{a_i}}_{ij}$). Here $\rho_{ij}$ is the density matrix $\rho$ reduced with respect to all  subsystems $a_k$ except $a_i$ and $a_j$:
$\rho_{ij}={\mbox{Tr}}_{\{a_k,\;k\neq i,j\}} \rho$ and superscript $T_{a_j}$ (or $T_{a_i}$)  means the transposition with respect to the subsystem $a_j$ (or $a_i$).
The choice of PPT criterion provides applicability of the represented algorithm to both pure and mixed states.}

The  results of this paper are based on the following definition.

{\bf Definition:} Let us consider the set of  bipartite entanglements of all
possible bipartite decompositions of the given quantum system. The minimal
entanglement out of this set  will be referred to as   the minimal entanglement
of bipartite decompositions (MEBD) of the given quantum system.

In accordance with this definition, 
the big value of MEBD means that the quantum system may not be decomposed into
two weakly entangled subsystems.

To clarify the importance of MEBD we refer to the  quantum register. Remember,
that the inherent property of quantum register is the simultaneous entanglement among all its
nodes. This entanglement allows one to reach  the exponential speedup of the quantum calculations in comparison with the classical ones. Thus, if the $N$-node quantum system may be separated into two weakly
entangled subsystems of $N_1$ and $N_2$ nodes ($N_1+N_2=N$), then the effective
volume of the register constructed on the basis of this quatum system is  either
$N_1$ or $N_2$ rather then $N$. By definition of MEBD we see that such
separation is impossible if only MEBD is  big. Thus,  MEBD  helps us, for
instance, to test 
whether all nodes of the quantum system may be effectively used as different bits in a quantum
register.

In accordance with the above  definition of MEBD, we  introduce a measure of
MEBD
 which vanishes if 
only there is at least one decomposition of the quantum system  into two weakly
entangled subsystems 
(note that the measure of  multipartite entanglement introduced in refs.
\cite{VPRK,WG} does not have this property). Namely,
let  the system $S_N$ have $M$ different decompositions into two
subsystems: 
\begin{eqnarray}\label{dec}
&&
S_N=A^{(1)}_i\cup A^{(2)}_i,\;\;A^{(1)}_i\cap
A^{(2)}_i=\emptyset,\;\;i=1,\dots,M,\\\nonumber
&&
M=\left\{
\begin{array}{ll}
\sum_{k=1}^{N/2-1}C_{N}^k+\frac{1}{2}C_N^{N/2}=
2^{N-1} -1,&N=2,4,6,\dots\cr
\sum_{k=1}^{(N-1)/2}C_{N}^k=
2^{N-1}-1, &N=3,5,7,\dots
\end{array}
\right. ,
\end{eqnarray}
{where $C^i_j$ are binomial coefficients: $\displaystyle C^i_j=\frac{j!}{i!(j-i)!}$.}
Then we  introduce the measure of MEBD  $E(S_N)$ by the following formula:
\begin{eqnarray}\label{SAB}\label{SAB_N}
 E({S_N}) = \min_{i=1,\dots,M} {\cal{N}}_{A^{(1)}_i,A^{(2)}_i}.
\end{eqnarray}
This formula is consistent with the definition of MEBD, because the measure  is
zero if only there are at least two  un-entangled subsystems $A^{(1)}_{i_0}$ and
$A^{(2)}_{i_0}$: 
${\cal{N}}_{A^{(1)}_{i_0},A^{(2)}_{i_0}}=0$. We consider that the value of MEBD
is big if its measure approaches unity, and the value of MEBD is small if  its
measure tends to zero.

Emphasize that the big  MEBD does not require the strong entanglement between
any two nodes of the quantum system. In other words, MEBD may be big   even if
some particular nodes are weakly entangled between each other. { This statement may be simply  justified using the property of the hierarchy of
double negativities \cite{VW}, which can be formulated as follows. 
Let the quantum system be separated into the set of subsystems (\ref{multipartite}).
Then
\begin{eqnarray}\label{HDN}
{\cal{N}}_{a_1,\{a_2,\dots,L\}} \ge {\cal{N}}_{a_1,\{a_2,\dots,L-1\}}\ge \cdots \ge {\cal{N}}_{a_1,a_2}.
\end{eqnarray}
Let us  apply this property to  } two simple examples of spin-1/2
systems with big MEBD where only some two-node double negativities are big. 
Hereafter we will  use notation $S_N=\{s_1,\dots,s_N\}$ for the $N$-node spin
system, where $s_i$ means the $i$th node.  

 First of all, we  turn to the three-node spin system  $S_3=\{s_1,s_2,s_3\}$,
which may be decomposed as follows:
\begin{eqnarray}\label{decN_3}
S_3=s_1 \cup \{s_2,s_3\} =s_2\cup \{s_1,s_3\}=s_3 \cup \{s_1,s_2\}.
\end{eqnarray}
In order to obtain big MEBD defined by Eq.(\ref{SAB})  it is enough to have two
big double negativities: ${\cal{N}}_{s_1,s_2}$ and
 ${\cal{N}}_{s_3,\{s_1,s_2\}}$.  In fact, using the property of the hierarchy of
double negativities (\ref{HDN}) we obtain that 
${\cal{N}}_{s_1,\{s_2,s_3\}}\ge {\cal{N}}_{s_1,s_2}$,
${\cal{N}}_{s_2,\{s_1,s_3\}}\ge {\cal{N}}_{s_2,s_1}={\cal{N}}_{s_1,s_2}$.
Thus, all ${\cal{N}}_{s_i,\{s_j,s_k\}}$   are big (all $i$, $j$ and $k$ are
different) and the system may not be decomposed into two weakly entangled
subsystems, i.e. MEBD of the system  $S_3$ is big. 

 The second example regards  the four node system $S_4=\{s_1,s_2,s_3,s_4\}$,
which may be decomposed into two subsystems as follows:
\begin{eqnarray}\label{ex_dec}\label{dec_4}
&&
S_4=s_1\cup \{s_2,s_3,s_4\}=
s_2\cup\{s_1,s_3,s_4\}=s_3\cup\{s_1,s_2,s_4\}=\\\nonumber
&&s_4\cup\{s_1,s_2,s_3\}= 
\{s_1,s_2\}\cup \{s_3,s_4\}=\{s_1,s_3\}\cup\{s_2,s_4\}=\\\nonumber
&&
\{s_1,s_4\}\cup\{s_2,s_3\}.
\end{eqnarray}
In order to achieve big MEBD, it is enough to have three big double
negativities:
 ${\cal{N}}_{s_1,s_2}$, ${\cal{N}}_{s_3,s_4}$ and
${\cal{N}}_{\{s_1,s_2\},\{s_3,s_4\}}$.  
Then, using the property of the hierarchy of double negativities (\ref{HDN}),
one can show  that the double negativities   for all decompositions 
(\ref{ex_dec}) are big.
In fact:
\begin{eqnarray}\label{neg}
&&
{\cal{N}}_{s_1,\{s_2,s_3,s_4\}}\ge
{\cal{N}}_{s_1,s_2},\;\;{\cal{N}}_{s_2,\{s_1,s_3,s_4\}}\ge
{\cal{N}}_{s_2,s_1}\equiv{\cal{N}}_{s_1,s_2} ,
\\\nonumber
&&{\cal{N}}_{s_3,\{s_1,s_2,s_4\}}\ge {\cal{N}}_{s_3,s_4} ,\;\;
{\cal{N}}_{s_4,\{s_1,s_2,s_3\}}\ge {\cal{N}}_{s_4,s_3}\equiv{\cal{N}}_{s_3,s_4},
\\\nonumber
&&{\cal{N}}_{\{s_1,s_3\},\{s_2,s_4\}}\ge {\cal{N}}_{s_1,s_2},
\;\;
{\cal{N}}_{\{s_1,s_4\},\{s_2,s_3\}}\ge {\cal{N}}_{s_1,s_2}.
\end{eqnarray}
 In result, { since  ${\cal{N}}_{s_1,s_2}$, ${\cal{N}}_{s_3,s_4}$ and
${\cal{N}}_{\{s_1,s_2\},\{s_3,s_4\}}$ are big by our assumption}, then the above  system $S_4$ may not be decomposed into two weakly
entangled subsystems, i.e. MEBD of the system  $S_4$ is big. 

\subsection{Lower estimation of MEBD}
\label{Section:low}
Note  that
 $ E(S_N)$ given by  formula (\ref{SAB_N})
 is cumbersome for calculations if $N$ is big, because $M$ increases very
quickly with the increase in $N$.
Thus, it is reasonable to introduce another quantity  giving a lower estimation
of MEBD in the system  $S_N$, which is based on the following remark.  

Let us fix some decomposition of $S_N$, say, $S_N=A^{(1)}_1\cup A^{(2)}_1$. {
Let $E({A^{(i)}_{1}})$, $i=1,2$, be measures of MEBD of subsystems $A^{(i)}_{1}$ (introduced in the previous decomposition), see Eq.(\ref{SAB}) with replacement $S_N$ by $A^{(i)}_{1}$.}
Then we state that if the quantity $E^{(1)}_1({S_N})$,  $E^{(1)}_1({S_N})=\min
(E({A^{(1)}_{1}}), E({A^{(2)}_{1}}), {\cal{N}}_{A^{(1)}_{1},A^{(2)}_{1}})$, is
big, then MEBD of  $S_N$ is big, i.e. $E(S_N)$ is big as well. The opposite is
not true in general, i.e. if  $E^{(1)}_1({S_N})$ is small, then MEBD of the
system may be big. This happens if there is another decomposition 
$S_N=A^{(1)}_2\cup A^{(2)}_2$ such that $E^{(1)}_2({S_N})=\min (E({A^{(1)}_{2}}),
E({A^{(2)}_{2}}), {\cal{N}}_{A^{(1)}_{2},A^{(2)}_{2}})$ is big. In other words we
may estimate MEBD of the system of $N$ nodes as follows.

Let us fix the $j$th  decomposition of $S_N$ into two subsystems,
$S_N=A^{(1)}_{j}\cup A^{(2)}_{j},\;\;1\le j\le M$,
and consider the following quantities:
\begin{eqnarray}\label{Eij}
E^{(1)}_j(S_N) = \min \left(E({A^{(1)}_{j}}),E({A^{(2)}_{j}}),
{\cal{N}}_{A^{(1)}_{j},A^{(2)}_{j}}\right),
\end{eqnarray}
{where $E({A^{(i)}_{j}})$ are  measures of MEBD of subsystems $A^{(i)}_{j}$.}
The manifold of all possible $E^{(1)}_j({S_N})$ will be referred to as 
${\cal{E}}^{(1)}({S_N})$:
${\cal{E}}^{(1)}({S_N})=\{E^{(1)}_j({S_N}),\;\;j=1,\dots,M \}$. It is
clear that the value $E^{(1)}({S_N})$, calculated by the  formula
{
\begin{eqnarray}\label{E11}
E^{(1)}({S_N})=\max\limits_{j=1,\dots,M}\; E^{(1)}_j({S_N}),
\end{eqnarray}
}
  satisfies the inequality ${{E}}^{(1)}({S_N})\le {{E}}({S_N})$  due to the
hierarchy of double negativities (\ref{HDN}).
Thus ${{E}}^{(1)}({S_N})$ may be referred to as the lower estimation of MEBD of
the system $S_N$. 

{We  may introduce another lower estimation as follows. Let us decompose each subsystem $A^{(1)}_{j}$ and $A^{(2)}_{j}$ into two
smaller ones:
\begin{eqnarray}\label{Dec21}
A^{(1)}_{j}=A^{(11)}_{jk}\cup
A^{(12)}_{jk},\;\;A^{(11)}_{jk}\cap
A^{(12)}_{jk}=\emptyset,\;\;1\le k\le M^{(1)}_{j}\\\label{Dec22}
A^{(2)}_{j}=A^{(21)}_{jn}\cup
A^{(22)}_{jn},\;\;A^{(21)}_{jn}\cap
A^{(22)}_{jn}=\emptyset,\;\;1\le n\le M^{(2)}_{j}.
\end{eqnarray} 
Here $M^{(l)}_{j}$, $l=1,2$, are the  numbers of all possible decompositions
(\ref{Dec21}) and (\ref{Dec22}) respectively  with fixed $j$.
Let $E^{(1)}_k({A^{(i)}_{j}})$ be lower estimation 
of measure of MEBD for the subsystem ${A^{(i)}_{j}}$, see Eq.(\ref{Eij}) for the definition of this estimation. Then another lower estimation may be introduced by the following formula:
\begin{eqnarray}
E^{(2)}_{jkn}({S_N}) =
 \min \left(E^{(1)}_k({A^{(1)}_{j}}),E^{(1)}_n({A^{(2)}_{j}}),
{\cal{N}}_{A^{(1)}_{j},A^{(2)}_{j}}\right).
\end{eqnarray}
}
The manifold of all possible $E^{(2)}_{jkn}({S_N})$  will be referred to as 
${\cal{E}}^{(2)}({S_N})$: ${\cal{E}}^{(2)}({S_N})=\{E^{(2)}_{jkn}({S_N}),\;\;
1\le j \le M,\; 
1\le k \le M^{(1)}_{j},\; 1\le n \le M^{(2)}_{j}\}$. We define  the
quantity  $E^{(2)}({S_N})$,
{
\begin{eqnarray}
E^{(2)}({S_N})=\max\limits_{j,k,n}\; E^{(2)}_{jkn}({S_N}).
\end{eqnarray}
}
 Again  one has  ${{E}}^{(2)}({S_N})\le {{E}}^{(1)}({S_N})\le {{E}}({S_N})$ due to
the hierarchy of double negativities (\ref{HDN}). Thus, the quantity ${{E}}^{(2)}({S_N})$ is
another lower estimation of MEBD  of the system  ${S_N}$.
This process may be continued involving further decompositions of the subsystems
$A^{(nm)}_{jk}$ into the smaller ones to find the manifolds
${\cal{E}}^{(p)}({S_N})$ and appropriate minima ${{E}}^{(p)}({S_N})$,
$p=3,4,\dots, J_{max}$,
where ${{E}}^{(J_{max})}({S_N})$ corresponds to the case when all subsystems
consist of two nodes.
 Again we obtain that
\begin{eqnarray}\label{hierarchy}
{ {E}}^{(J_{max})}({S_N}) \le { {E}}^{(J_{max}-1)}({S_N})\le \cdots \le {
{E}}^{(1)}({S_N})\le E(S_N)
\end{eqnarray}
due to the hierarchy of double negativities (\ref{HDN}).

All in all, if there is a big element 
 $e_0\in {\cal{E}}^{(k)}({S_N})$, then we may write  $ e_0\le  E^{(k)}({S_N}) \le
E(S_N)$, i.e. $e_0$ may serve as the lower estimation of MEBD of our system.
{ This low estimation will be found in examples of  Sec.\ref{Section:Hdz}.}

{
It is worthwhile to clarify, whether the values  $E^{(k)}(S_N)$ can be used for 
the reasonable approximation of
$E(S_N)$ or their are just lower bounds of MEBD.  
First of all, it is necessary to remark that $E^{(k)}(S_N)$ is significantly less then $E(S_N)$, which is a consequence of the additivity property of the entanglement \cite{B}. On the other hand, this additivity property allows us to  improve estimation as follows. For simplicity, we consider  the  case  $k=1$ and symmetrical decomposition $S_N=A^{(1)}_1 \cup A^{(1)}_1$. 
Of course we  may not apply the additivity property to our case directly because, for instance,   $\rho\neq \rho^{A^{(1)}_1} \otimes \rho^{A^{(1)}_1}$. Nevertheless we use it  for the rough estimations. Assume that ${\cal{N}}_{A^{(1)}_1,A^{(2)}_1}>E(A^{(1)}_1)$, which is reasonable for the 
big MEBD when $E(S_N)\sim 1$. Then $E^{(1)}_1(S_N)=E(A^{(1)}_1)$.
In accordance with the additivity property
we may write
\begin{eqnarray}\label{lowappr}
 E(S_N) \sim 2 E(A^{(1)}_1) \;\;\Rightarrow  E(A^{(1)}_1)\sim \frac{1}{2} E(S_N).
\end{eqnarray} 
 Consequently we may propose that  $E^{(1)}_1(S_N)$   (and $E^{(1)}(S_N)$) is big if  $E^{(1)}_1(S_N)\approx 1/2$. In this case  $E(S_N) \sim 1$. Thus $E^{(1)}_1(S_N)$  may be used to approximate  $E(S_N)$.
 }


\section{The dynamics of  MEBD in the homogeneous spin-1/2 chains governed by
the $H_{dz}$ Hamiltonian.}
\label{Section:Hdz}
{
The dynamics of entanglement significantly depends on the  physical scenario, in particular, on the initial state of the quantum system. 
For instance, the dissipative dinamics of multipartite entanglement is considered in set of papers, see \cite{MCKB,CGPEK}. In this section we consider the quasiperiodic dynamics of MEBD  caused by the special initial condition in the spin-1/2 homogeneous chain governed by the  $H_{dz}$ Hamiltonian}
\begin{eqnarray}\label{Hdz}
 {\cal{H}}_{dz} &=&
\sum_{{i,j=1}\atop{j>i}}^{N}
D_{i,j}(I_{i,x}I_{j,x} + I_{i,y}I_{j,y}-2  I_{i,z}I_{j,z}),\;\;
D_{i,j}=\frac{\gamma^2 \hbar}{r_{ij}^3}
\end{eqnarray}
{ in the strong external magnetic field.}
{Here $\gamma$ is the gyromagnetic ratio, $r_{ij}$ are distances between the $i$th and the $j$th nodes, $\hbar $ is the Planck constant.  Hamiltonian $H_{dz}$ is the secular part of the Hamiltonian describing the dipole-dipole interaction in the strong external magnetic field \cite{G}. }
This Hamiltonian is known to commute with $I_z$ (the $z$-projection of the total
spin). {
The density matrix,  obtained as the solution to the Liouville evolution equation 
$i \rho_t = [{\cal{H}}_{dz},\rho]$ ($\hbar=1$),
reads}
\begin{eqnarray}\label{rhot}
\rho(t)=e^{-i {\cal{H}}_{dz}t} \rho_0 e^{i {\cal{H}}_{dz}t},
\end{eqnarray}
where $\rho_0$ is the initial density matrix. 
In the homogeneous chain  $D_{i,i+1}\equiv D$,
$i=1,\dots,N-1$. Hereafter we use the dimensionless time $\tau=D t$. It is pointed above, that the dynamics of entanglement (and, consequently, the dynamics of MEBD) strongly depends on  the initial state of the
spin chain, in particular, on   the number of initially exited spins {(i.e. the spins which are directed opposite to the external magnetic fields)}. {
We use notation $|\Psi_0(S_N)\rangle $ for the initial wave function, so that $\rho_0(S_N)=|\Psi_0(S_N)\rangle \langle \Psi_0(S_N)|$.}
We  have found that in order to achieve big MEBD at some time moment in the
homogeneous spin chains with $N=3,4,6,8$ one has to take the initial condition {$|\Psi_0(S_N)\rangle $}
with one, two, three and four exited spins respectively.
We consider that MEBD is big { at some time moment $\tau_N$} if the introduced measure $E(S_N)$ (see
Eq.(\ref{SAB})) approaches unity {at this time moment}. Emphasize, that the problem of the strong
simultaneous entanglement between any two nodes  is not resolved even for the
spin chains of several nodes. For this reason the observation of big MEBD in
the short chains is valuable. It is found that a big MEBD may be  achieved even
though the entanglement between two particular nodes is weak. In all examples
given below, $\tau_N$, $N=3,4,6,8$, means the 
time moment when  big MEBD is achieved for the first time { in the homogeneous $N$-node spin chain}. 
{
Hereafter we do not write $\tau$ as argument of functions for the sake of simplicity. We use the Dirac notations $|n_1\dots n_N\rangle$ for the wave functions, where $n_i=0$  if the $i$th spin is directed along the external magnetic field,  or $n_i=1$  if the $i$th spin is directed opposite to  the external magnetic field, i.e. the $i$th spin is excited.
}

\begin{figure*}
   \epsfig{file=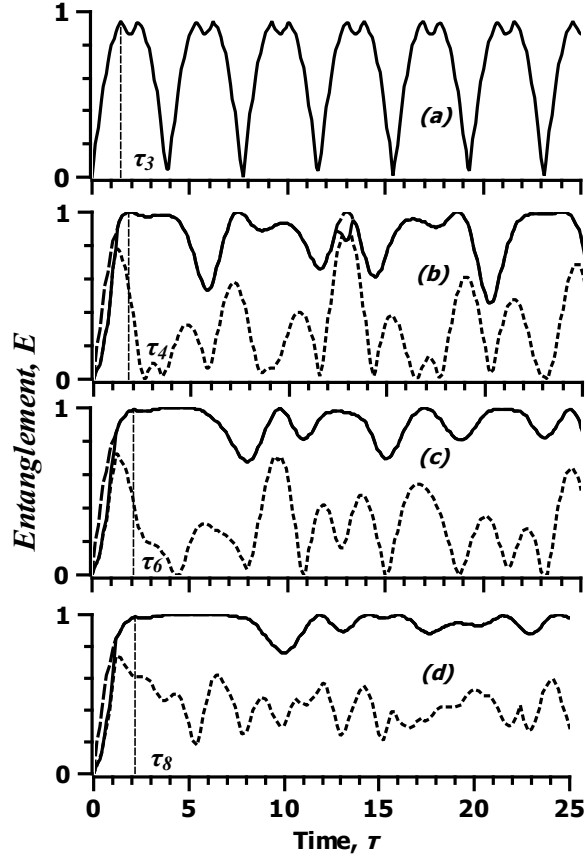
   ,scale=0.8,angle=270}
  \caption{The evolution of the entanglement $E(S_N)$ (solid lines), of the lower
estimation of the entanglement  $E^{(1)}_1({S_N})$ (dotted lines) and of  $\tilde E(S_N)=\min_i({\cal{N}}_{s_i,rest^N_i})$ (dashed lines) in  chains $S_N$  with the
special initial conditions $|\Psi_{0}({S_N})\rangle$, $N=3,4,6,8$: 
$(a)$  $N=3$, $|\Psi_{0}({S_3})\rangle=|010\rangle$;
$(b)$ $N=4$, $|\Psi_{0}({S_4})\rangle=|1001\rangle$;
$(c)$ $N=6$, $|\Psi_{0}({S_6})\rangle=|100110\rangle$;
$(d)$ $N=8$, $|\Psi_{0}({S_8})\rangle=|10011001\rangle$}
  \label{Fig:N_3} 
\end{figure*}

{\bf Example 1: $N=3$.}
\label{Section:N3}
As a simple example we consider the chain of three spins, $S_3$,
 with the initial state
$|\Psi_0(S_3)\rangle=|010\rangle $. 
 All possible decompositions of $S_3$ into two subsystems  are given in
Eq.(\ref{decN_3}). 
The evolution of  
$E({S_3})=\min({\cal{N}}_{s_1,\{s_2,s_3\}},{\cal{N}}_{s_2,\{s_1,s_3\}
},
{\cal{N}}_{s_3,\{s_1,s_2\}})$ (see Eq.(\ref{SAB})) is shown in
Fig.\ref{Fig:N_3}$(a)$.
The first maximum of this function  is achieved at $\tau_3= 1.505$,
$E({S_3})=0.943$, i.e. MEBD is big at $\tau_3$.

{\bf Example 2: $N=4$.}
\label{Section:N4}
Consider the chain of four spins,
$S_4$, with the initial state
$|\Psi_{0}({S_4})\rangle=|1001\rangle$. 
In this case
all decompositions of $S_4$ into two subsystems are  given in 
Eq.(\ref{ex_dec}).
Note, that  the double negativities
${\cal{N}}_{s_{j},\{s_{k},s_{i_3},s_{i_4}\}}$ 
(all $j,k,i_3,i_4$ are different) are less then the double negativities
associated with other decompositions in  set (\ref{ex_dec}), except the short interval in the vicinity of $\tau=0$.  
Using  definition (\ref{SAB}) we calculate { the evolution of}  $E({S_4})$ which 
is shown in Fig.\ref{Fig:N_3}$(b)$.
The first maximum of this function  is achieved at $\tau_4= 1.819$,
$E({S_4})=1.000$, i.e. MEBD is big at $\tau_4$ {so that the system may not be decomposed into two weakly entangled subsystem at the time moment $\tau_4$.}

{Let us calculate the lower estimation of  MEBD using the technique developed in Sec.\ref{Section:low}. Let $A^{(1)}_1=\{s_1,s_2\}$, $A^{(2)}_1=\{s_3,s_4\}$ and calculate 
$E^{1}_1({S_4})=\min(E({A^{(1)}_1}),E({A^{(2)}_1}),{\cal{N}}_{A^{(1)}_1,A^{
(2)}_1}) \equiv \min({\cal{N}}_{s_1,s_2},{\cal{N}}_{s_3,s_4},{\cal{N}}_{\{s_1,s_2\},{\{s_3,s_4\}}})$. The evolution of this function is shown in Fig.\ref{Fig:N_3}$(b)$), dotted line.}

{\bf Example 3: $N=6$.}
\label{Section:N6}
Consider the chain of six spins $S_6$ with the initial state
$|\Psi_0(S_6)\rangle=|100110\rangle$. 
The system may be decomposed into two subsystems as follows:
\begin{eqnarray}\label{decN_6}
S_6&=&s_{i} \cup \{s_{j},s_{k},s_{l},s_{n},s_{m}\}=
\{s_{i},s_{j}\} \cup \{s_{k},s_{l},s_{n},s_{m}\}=
\\\nonumber
&&
\{s_{i},s_{j},s_{k}\}\cup \{s_{l},s_{n},s_{m}\} ,\\\nonumber
&&
{\mbox{all $i,j,k,l,n,m$ are different}},\;\;i,j,k,l,n,m=1,\dots,6.
\end{eqnarray}
Note, that the double negativities
${\cal{N}}_{s_{i},\{s_{j},s_{k},s_{l},s_{n},s_{m}\}}$ (all
$i,j,k,l,n,m$ are different)
are less then the double negativities associated with other decompositions in 
set (\ref{decN_6}), except the short interval in the vicinity of $\tau=0$.
 Using definition (\ref{SAB}) we calculate {the evolution of} $E({S_6})$ which is
shown in Fig.\ref{Fig:N_3}$(c)$.
The first maximum  of this function (i.e. a big value of  MEBD) is achieved at $\tau_6= 2.110$,
$E({S_6})=0.992$.

{
Let us find the lower estimation of  MEBD.
Let $A^{(1)}_1=\{s_1,s_2,s_3,s_4\}$, $A^{(2)}_1=\{s_5,s_6\}$ and calculate 
$E^{1}_1({S_6})=\min(E({A^{(1)}_1}),E({A^{(2)}_1}),{\cal{N}}_{A^{(1)}_1,A^{
(2)}_1})$. Here $E({A^{(1)}_1})$ is given by eq.(\ref{SAB}) with decompositions of $A^{(1)}_1$ given by eq.(\ref{ex_dec}), while  $E({A^{(2)}_1})\equiv {\cal{N}}_{5,6}$. The evolution of $E^{1}_1({S_6})$ is shown in Fig.\ref{Fig:N_3}$(c)$, dotted line.
}

{\bf Example 4: $N=8$.}
\label{Section:N8}
{Consider the spin chain $S_8$
with the initial state
$|\Psi_0(S_8)\rangle=|10011001\rangle$. 
The system may be decomposed into two subsystems as follows:
\begin{eqnarray}\label{decN_8}
S_6&=&s_{i} \cup \{s_{j},s_{k},s_{l},s_{n},s_{m},s_{p},s_{q}\}=
\{s_{i},s_{j}\} \cup \{s_{k},s_{l},s_{n},s_{m},s_{p},s_{q}\}=
\\\nonumber
&&
\{s_{i},s_{j},s_{k}\}\cup \{s_{l},s_{n},s_{m},s_{p},s_{q}\} =\{s_{i},s_{j},s_{k},s_{l}\}\cup \{s_{n},s_{m},s_{p},s_{q}\}, \\\nonumber
&&
{\mbox{all $i,j,k,l,n,m,p,q$ are different}},\;\;i,j,k,l,n,m,p,q=1,\dots,8.
\end{eqnarray}
Note, that the double negativities
${\cal{N}}_{s_{i},\{s_{j},s_{k},s_{l},s_{n},s_{m},s_{p},s_{q}\}}$ (all
$i,j,k,l,n,m,p,q$ are different)
are less then the double negativities associated with other decompositions in 
set (\ref{decN_8}), except the short interval in the vicinity of $\tau=0$.
 Using definition (\ref{SAB}) we calculate {the evolution of} $E({S_8})$ which is
shown in Fig.\ref{Fig:N_3}$(d)$.
The first maximum of this function is achieved at $\tau_8= 2.193$,
$E({S_8})=0.988$, i.e. MEBD is big at $\tau_8$.
}

We also calculate
$E^{1}_1({S_8})$ (see Eq.(\ref{Eij})) with $A^{(1)}_1=\{s_1,s_2,s_3,s_4\}$ and
$A^{(2)}_1=\{s_5,s_6,s_7,s_8\}$.
{ The evolution of }
$E^{1}_1({S_8})=\min(E({A^{(1)}_1}),E({A^{(2)}_1}),{\cal{N}}_{A^{(1)}_1,A^{
(2)}_1}) $ is shown in Fig.\ref{Fig:N_3}$(d)$, dotted line. 
Here $E({A^{(1)}_1})$ and $E({A^{(2)}_1})$ are minima of all double negativities
associated with all possible bipartite decompositions of $A^{(1)}_1$ and
$A^{(2)}_1$, where the
decompositions of $A^{(1)}_1$ coincide with those given in Eq.(\ref{dec_4}), while
the decompositions of $A^{(2)}_1$ are given in Eq.(\ref{dec_4}) with replacements
$s_i\to s_{i+4}$, $i=1,2,3,4$. 

{ Evolutions of MEBD is shown in Fig.\ref{Fig:N_3} (solid lines) demonstrates that the graphs of  MEBD for systems with $N>3$  oscillate around some value in the vicinity of unity. The amplitude of these oscillations decreases with the increase in $N$ so that MEBD tends to straight line $E=1$. Regarding the lower estimations (dotted lines), one must note that they do not approach unit. They also oscillate and the   amplitudes of these oscillations decrease with the increase in $N$.  The low estimation do never approach zero for $N=8$ and $\tau>0$. It may be considered as a function oscillating around some value $E_l\approx 0.4$. Comparison of solids and dotted lines shows that the maxima of $E(S_N)$ and $E^{(1)}_1(S_N)$ well correlate with each other, although the maxima of $E(S_N)$ are less localized then those of $E^{(1)}_1(S_N)$. The fact that the low estimation is  less then   MEBD  is a  consequence of the fact that MEBD is  a measure of the "collective" entanglement which is carried by the dencity matrix $\rho$. At the same time,  calculating the low estimation, we use the reduced dencity matrices  which do not contain the complete   information about the "collective" entanglement. This is in agreement with the estimation given in eq.(\ref{lowappr}).

It is also important to note, that the evolution of MEBD shown in Fig.\ref{Fig:N_3}  almost coinsides with the evolution of the quantity  $\tilde E(S_N)=\min\limits_i({\cal{N}}_{s_i,rest^N_i})$ (dashed lines), where $rest^N_i$ means the whole  system $S_N$ except the node $s_i$ (solid and dashed lines coinside almost  everywhere). If so then one can use $\tilde E(S_N)$ in order to estimate MEBD with high precision. However, this statement remains unproved for an arbitrary quantum system.}

The first maxima of the functions $E(S_N)$ together with the
appropriate time moments 
are collected in  Table 1. 
{ It is important to realize how long are the found time intervals $\tau_N$. These intervals must be short enough in order to provide the coherent manipulations by qbits in a quantum circuit.
For this purpose we note that the dynamics of MEBD is quasiperiodic, similar to the dynamics of 
the quantum state transfer probability along the spin chain. Since both MEBD and quantum state transfer deal with the same physical object (spin chain) it is reasonable to 
 compare the time intervals $\tau_N$ (obtained in our paper)  with 
the time intervals required for  the end-to-end quantum state transfer. This quantum process is well studied 
\cite{CDEL,KS,VGIZ,GMT,DZ}  and the minimal possible time interval is established \cite{CDEL}: it equals to $\pi$ in dimensionless variables. In our case, $\tau_N< \pi$, which 
indicates that the big values of MEBD are achievable during the reasonable time intervals. 
}



 \begin{table*}[!htb]
\begin{tabular}{|p{1.4cm}|p{1cm}|p{1cm}|p{1cm} |p{1cm}|}
\hline
$N$&3&4&6&8 \\\hline
$E(S_N)$&0.943&1.000&0.992&0.988 \\
$\tau_N$&1.505&1.819&2.110& 2.193
\\\hline
\end{tabular}
\label{Table:HPST4}
\caption{The maximal values $E(S_N)$,
$N=3,4,6,8$,
and the appropriate time moments $\tau_N$
}
\end{table*}

\section{Conclusions.}
\label{Section:conclusions}

We introduce  MEBD as a witness of the strong entanglement in a quantum system.
Although the strong entanglement between any two nodes of a quantum system is
hardly achievable simultaneously, the big MEBD is quite realizable.
The basic feature of MEBD is that its measure  vanishes if only there is at
least one  decomposition of the quantum system  into two weakly entangled
subsystems. This  feature does not appear in other measures of entanglement. 
Having this property, MEBD may be used, for instance, to test whether $N$-node
spin system may be a candidate for the $N$-qbit quantum register or this quantum
system must be separated into two registers of smaller size. 

 We suggest  a method of lower estimation of MEBD which is useful for the large
systems.

 The dynamics of double negativities in the homogeneous  spin-1/2  chains governed by
the $H_{dz}$ Hamiltonian with the appropriate initial conditions
 demonstrates us a possibility to obtain 
big MEBD of a quantum system {during the relatively short time intervals $\tau_N<\pi$}, { which is comparable with the shortest time intervals obtained for the quantum state transfer along the spin chains.}
 We consider only the short spin chains because the
problem of strong entanglement between all nodes  is unresolved even for small
quantum systems. 

Author thanks Professor  E.B.Fel'dman for useful discussions.
This work is supported 
by the Program of the Presidium of RAS 
No.7 "Development of methods of obtaining chemical compounds and creation of new
materials".


\begin{thebibliography}{99}

{

\bibitem{NC}
 M.A.Nielsen, I.L.Chuang, Quantum Computation and Quantum Information (Cambridge University Press, 
Cambridge, 2000)

\bibitem{Lea}
D.Leibfried {\it et al.}, Nature, {\bf 422} (2003) 412

\bibitem{YPANT}
T.Yamamoto, Yu.A.Pashkin, O.Astafiev, Y.Nakamura and J.S.Tsai, Nature, {\bf 425} (2003) 941

\bibitem{SDKMRM}
D.Schrader, I. Dotsenko, M. Khudaverdyan, Y. Miroshnychenko, A. Rauschenbeutel and D. Meschede, Phys. Rev. Lett. {\bf 93}  (2004) 150501

\bibitem{DB}
W.D\"ur and H.J.Briegel, Phys.Rev.Lett. {\bf 90} (2003) 067901

\bibitem{LBBKK}
 Y.L.Lim, S.D.Barrett, A.Beige1, P.Kok, and L.C.Kwek, Phys. Rev. A {\bf 73} (2006) 012304 

\bibitem{ODH}
D.K.L.Oi, S.J.Devitt1 and L.C.L.Hollenberg,
Phys. Rev. A {\bf 74} (2006) 052313 
}

\bibitem{AFOV}
L.Amico, R.Fazio, A.Osterloh and V.Ventral, Rev. Mod. Phys. {\bf 80}, 517 
(2008)

\bibitem{HHHH}
R.Horodecki,
P.Horodecki, M.Horodecki and K.Horodecki, Rev. Mod. Phys. {\bf 81}, 865  (2009)


\bibitem{HW}
S.Hill and W.K.Wootters, Phys. Rev. Lett. {\bf 78}, 5022 (1997)

\bibitem{W}
W.K.Wootters, Phys. Rev. Lett. {\bf 80}, 2245 (1998)

\bibitem{P}
A.Peres, Phys. Rev. Lett. {\bf 77}, 1413 (1996)

\bibitem{VW}
G.Vidal and R.F.Werner, Phys. Rev. A {\bf 65}, 032314 (2002)

\bibitem{BDSW} 
C.H.Bennett, D.P.DiVincenzo, J.Sm olin and W.K.Wootters, Phys. Rev. A {\bf 54},
3824 (1996)

\bibitem{VPRK}
V.Ventral, M.B.Plenio, M.A.Rippin and P.L.Knight,
Phys. Rev. Lett. {\bf 78}, 2275 (1997)

\bibitem{WG}
T.C.Wei and P.M.Goldbart, Phys. Rev. A {\bf 68}, 042307 (2003)



\bibitem{CDEL}
M.Christandl, N.Datta, A.Ekert and A.J.Landahl, Phys.Rev.Lett. {\bf 92}, 187902 (2004)

\bibitem{KS}
P.Karbach and J.Stolze, Phys.Rev.A, {\bf 72}, 030301(R) (2005)



\bibitem{VGIZ}
L.Campos Venuti, S.M.Giampaolo, F.Illuminati and P.Zanardi,
Phys. Rev. A {\bf 76}, 052328 (2007) 



\bibitem{GMT}
G.Gualdi, I.Marzoli and P.Tombesi,
New J. Phys. {\bf 11}, 063038 (2009)



\bibitem{DZ} S.I.Doronin, A.I.Zenchuk, Phys. Rev. A {\bf 81}, 022321 (2010)  

{
\bibitem{MCKB}
F.Mintert, A.R.R.Carvalho, M.Ku\'s and A.Buchleitner, 
Physics Reports {\bf 415} (2005) 207

\bibitem{CGPEK}
 K.S.Choi, A.Goban, S.B.Papp, S.J. van Enk and H.J.Kimble
    Nature {\bf 468} (2010) 412

\bibitem{B}
D. Bru$\beta$,  J. Math. Phys. 43 (2002) 4237
}

\bibitem{G}
M.Goldman, Spin Temperature and Nuclear Magnetic Resonance in Solids (Clarendon, Oxford, 1970)





\end{thebibliography}
\end{document}